\documentclass[traditabstract]{aa}
\usepackage{txfonts,amssymb,graphicx}
\usepackage{natbib}
\newcommand{\tbl}[1]{\mbox{Table\hspace{0.3em}\ref{#1}}}
\newcommand{\fig}[1]{\mbox{Figure\hspace{0.2em}\ref{#1}}}
\begin{document}
\title{An Oort cloud origin of the Halley-type comets}
\author{J.-H. Wang \and R. Brasser}
\institute{Institute for Astronomy and Astrophysics, Academia Sinica; 11F AS/NTU building, 1 Roosevelt Rd., Sec. 4, Taipei 10617,
Taiwan}
\date{}
\abstract{The origin of the Halley-type comets (HTCs) is one of the last mysteries of the dynamical evolution of the Solar System.
Prior investigation into their origin has focused on two source regions: the Oort cloud and the Scattered Disc. From the former it
has been difficult to reproduce the non-isotropic, prograde skew in the inclination distribution of the observed HTCs without
invoking a multi-component Oort cloud model and specific fading of the comets. The Scattered Disc origin fares better but suffers from
needing an order of magnitude more mass than is currently advocated by theory and observations. Here we revisit the Oort cloud origin
and include cometary fading. Our observational sample stems from the JPL catalogue. We only keep comets discovered and observed after
1950  but place no a priori restriction on the maximum perihelion distance of observational completeness. We then numerically evolve
half a million comets from the Oort cloud through the realm of the giant planets and keep track of their number of perihelion passages
with perihelion distance $q<2.5$~AU, below which the activity is supposed to increase considerably. We can simultaneously fit the HTC
inclination and semi-major axis distribution very well with a power law fading function of the form $m^{-k}$, where $m$ is the number
of perihelion passages with $q<2.5$~AU and $k$ is the fading index. We match both the inclination and semi-major axis distributions
when $k \sim 1$ and the maximum imposed perihelion distance of the observed sample is $q_{\rm m} \sim 1.8$~AU. The value of $k$ is
higher than the one obtained for the Long-Period Comets (LPCs), for which typically $k \sim 0.7$. This increase in $k$ is most likely
the result of cometary surface processes. We argue the HTC sample is now most likely complete for $q_{\rm m}<1.8$~AU. We calculate that
the steady-state number of active HTCs with diameter $D>2.3$~km and $q<1.8$~AU is of the order of 100.}
\keywords{fill in when accepted}
\titlerunning{}
\maketitle

\section{Introduction}
The Solar System is host to a large population of comets, which tend to be concentrated in two components, i.e. the Oort cloud and
Scattered Disc \citep{1997Sci...276.1670D}. When examining the orbital data of new comets entering the inner solar system,
\cite{1950BAN....11...91O} noted that the distribution of reciprocal semi-major axis of these comets showed a sharp increase when
$1/a<5 \times 10^{-4}$~AU$^{-1}$. The observed semi-major axis distribution led Oort to suggest that the Sun is surrounded by a cloud
of comets in the region between 20\,000~AU to 150\,000~AU, and that it contains approximately $10^{11}$ comets with isotropic
inclination and random perihelia. Oort showed that the action of passing stars is able to account for the isotropic inclination
distribution, random perihelia, the continuous injection of new comets on trajectories that lead to perihelia close enough to the Sun
to become visible, and the fact that most new comets have near-parabolic orbits. This hypothesized cloud of comets surrounding the Sun
is now called the 'Oort cloud' (OC). \\
The only method with which we can infer the number of comets in the OC is by determining the flux of long-period comets (LPCs) and
Oort cloud comets (OCCs). The former are comets with a period longer than 200~yr that have had multiple passages through the solar
system while the latter are comets that are having their first perihelion passage in the inner solar system.\\
For a new comet to enter the planetary region from the OC, it must have sufficiently low angular momentum to allow it to
penetrate to small perihelion distances. For a given comet, its angular momentum per unit mass is given by $v_{\rm t} r$, where 
$v_{\rm t}$ is its transverse velocity and $r$ is the distance from the Sun. In the OC, where $r$ is large, $v_{\rm t}$ must be very
low to allow the comet to eventually come close to the Sun. Oort (1950) recognized this and added that the maximum allowable $v_{\rm
t}$ for a given distance $r$ would be a function of $r$. He gave this the name `loss cone' because comets within the velocity cone
could enter the planetary region and be ejected or captured to a shorter period orbit by the perturbations of the giant planets,
principally Jupiter and Saturn (which would require a perihelion distance $q<15$ AU). In other words, the comets would be `lost' from
the OC. These `lost' comets remain in the loss cone until ejected or destroyed. Thus the presence of Jupiter and Saturn presents an
obstacle for new comets to overcome in order to enter the inner solar system. This is often called the Jupiter-Saturn barrier.\\
There exist two agents which perturb the comets in the cloud onto orbits that enter the inner solar system: passing stars 
\citep{1980A&A....85..191W, 1981AJ.....86.1730H} and the Galactic tide \citep{1986Icar...65...13H, Levison2001}. 
The passing stars cause usually small random deviations in the orbital energy and other orbital elements. The Galactic tide on the
other hand systematically modifies the angular momentum of the comets at constant orbital energy. \cite{1986Icar...65...13H} discovered
that if the semi-major axis of the comet is large enough then the comet's change in perihelion can exceed 10~AU in a single orbit and
'jump' across the orbits of Jupiter and Saturn and thus not suffer their perturbations before becoming visible; in this particular case
the comet is considered a new comet. However, the \cite{1986Icar...65...13H} approximation of the Galactic tide only used the vertical
component, which is an order of magnitude stronger than the radial components. In this approximation the $z$-component of the comet's
orbital angular momentum in the Galactic plane is conserved and the comets follow closed trajectories in the $q-\omega$ plane (with $q$
being the perihelion distance and $\omega$ being the argument of perihelion). Including the radial tides breaks this conservation and
the flux of comets to the inner solar system from the OC is increased \citep{Levison2006}. The trajectories that lead comets into the
inner solar system should be quickly depleted, were it not for the passing stars to refill them \citep{2008CeMDA.102..111R}. The
synergy between these two perturbing agents ensures there is a roughly steady supply of comets entering the inner solar system. The
story is slightly more complicated because not all comets need to jump over the Jupiter-Saturn barrier: a small subset is able to sneak
through and undergo small enough energy changes that they still formally reside in the cloud \citep{2010MNRAS.404.1886K,
2013Icar..222...20F}, but the main argument still holds. \\
The OCCs form just a subset of comets that are visible from the Oort cloud. The OCCs are generally thought to be on their first
passage through the inner solar system, but may have had a few passages beyond $\sim$5~AU. However, we also have observational
evidence of comets that are dynamically evolved. These are the LPCs: they have undergone many passages through the realm of the giant
planets and/or the inner solar system and most of them are dynamically decoupled from the Oort cloud. A subset of these are the
Halley-type comets, or HTCs, named after its progenitor, comet 1P/Halley.\\
The second reservoir of cometary objects is the Scattered Disc \citep{1997Sci...276.1670D}. This source population resides on dynamically
unstable orbits beyond the orbit of Neptune. These bodies dynamically interact with Neptune and reside generally on low-inclination
orbits. The Scattered Disc is the most likely source of the so-called Jupiter-family comets (JFCs) \citep{1997Sci...276.1670D, 
2008ApJ...687..714V, 2013Icar..225...40B}.\\
Visible comets tend to historically be categorized into several groups: the JFCs, HTCs, LPCs and OCCs. Here we adhere to the
definitions of \cite{1996ASPC..107..173L} with a few modifications. These are:
\begin{itemize}
\item JFCs: $T_{\rm J} \in (2,3)$ and $a<7.35$~AU (Period $P<20$~yr),
\item HTCs: $T_{\rm J} <2$, $q>0.01$~AU and $a < 34.2$~AU (Period $P<200$~yr),
\item LPCs: $T_{\rm J}<2$, $q>0.01$~AU and  $34.2<a<10\,000$~AU (Period $P>200$~yr),
\item OCCs: $q>0.01$~AU and $a>10\,000$~AU.
\end{itemize}
Here $a$ is the semi-major axis, $P$ is the period of a comet, and $T_{\rm J}$ is the Tisserand parameter with respect to Jupiter. The
criterion $q>0.01$~AU was applied in order to exclude the Sun-grazing Kreutz comets. Dormant comets were also excluded in our work.
We realize that the above populations are not all inclusive: we did not include the Encke type and the Chiron type. There is no
definition for comets with $T_{\rm J}>2$ and $P>20$~yr, though these are generally included in the Centaur population.\\
The origin of the HTCs has long been controversial. The best attempts to determine the source of the HTCs were done by
\cite{1994A&A...281..911F}, \cite{Levison2001}, \cite{2002MNRAS.333..835N}, and \cite{Levison2006}.\\
\cite{1994A&A...281..911F} investigated the dynamical behaviour of LPCs coming from the Oort cloud and the role of perturbations by
the planets. They used a combination of a Monte Carlo random walk method and \"{O}pik's equations \citep{1951PRIA...54..165O} and
discovered that an initially isotropic distribution may result in a flattened distribution of evolved comets because of  the
inclination dependence on planetary energy kicks and the tendency of near-perpendicular comets to become retrograde. They find that the
typical comet undergoes 400 revolutions before becoming extinct, and combined with their capture probability they suggest there are of
the order of 300 active HTCs.\\
\cite{Levison2001} studied an Oort cloud origin for the HTCs but concluded that the non-isotropy of the HTCs could only be reproduced
with an Oort cloud that had a non-isotropic structure: it had to be flattened in the inner parts and isotropic in its outer parts.
\cite{Levison2001} included cometary fading in their simulations but they found that this alone was not enough to match the HTC
inclination distribution. \cite{2002MNRAS.333..835N} reached a different conclusion based on results of their Monte Carlo simulations:
the outer Oort cloud could be the source of the HTCs and their number and inclination distribution could be fairly well reproduced when
cometary fading and disruption was taken into account. However, when adding the inner Oort cloud they produced far too many HTCs,
which \cite{Levison2001} suggested was needed in their model. Unfortunately \cite{2002MNRAS.333..835N} did not try to match the 
semi-major axis distribution, something that Levison et al. (2001) did try to do. In addition, the initial conditions of both of these
works were artificial because they started the comets on planet-crossing orbits.\\
These issues led \cite{Levison2006} to suggest that the HTCs originate from the outer edge of the Scattered Disc. This latter reservoir
has a flattened inclination distribution \citep{1997Icar..127...13L} and the effects of the Galactic tide at the outer edge is able to
supply the retrograde HTCs. Their simulations match the inclination and semi-major axis distribution of the HTCs reasonably well. The
problem with this scenario is that it requires the Scattered Disc to be more than an order of magnitude more massive than most
estimates thus far \citep{Gomes2008, 2013Icar..225...40B}. This leads us back to the Oort cloud being the most likely possible
source.\\
Here we aim to demonstrate that the Oort cloud is the most likely source of the HTCs and that a single dynamical model with
cometary fading is sufficient to reproduce their current inclination and semi-major axis distributions. This paper is organized as
follows.\\
In the next section we present the observational dataset that we use for comparison. Section~3 details our numerical methods. In
Section~4 we present the results of our simulations and comparison with the observed dataset followed by a short discussion in
Section~5. We present a summary and conclusions in the last section.

\section{Observational Dataset}
In this study we determine whether or not the Oort cloud is the major source of the HTCs. Even though the Scattered Disc and Kuiper
belt do generate some HTCs \citep{1997Icar..127...13L}, the production probability is low and their orbital distribution is
inconsistent with the observed one. Thus, in this work we shall not consider the Scattered Disc and Kuiper belt as a source.\\
To compare our simulation with the observational data we need to have an observational catalogue of comets  that is as complete as
possible. There exist two easily accessible sources of cometary catalogues: the Catalogue of Cometary Orbits 2008 \citep{Marsden2008}
and the JPL Small-Body Search Engine\footnote{http://ssd.jpl.nasa.gov/sbdb\_query.cgi}.\\
The main difference between these two catalogues is that \cite{Marsden2008} did a backward integration and evaluated the orbital
elements when the comets were far away from the planetary region. They thus obtain the original orbital elements. On the other
hand, JPL evaluates the orbital elements when the comets are still in the planetary region. At present the Catalogue of Cometary
Orbits is no longer being updated (G. Williams, personal communication) and therefore it does not contain many recent cometary
discoveries from surveys such as WISE, Pan-STARRS and Catalina. These new comets are listed in the JPL catalogue, however. Therefore, in our
work we decided to make use of the JPL catalogue. As of July 2013, based on the criteria mentioned in previous section, the JPL
catalogue contains 427 LPCs and 60 HTCs. All HTCs based on our definition are tabulated in \tbl{tab:jplhtc}. One thing to note in the
table is that all these comets were discovered at various years, some even dating back to the 18th century.

\begin{figure}[ht]
\resizebox{\hsize}{!}{\includegraphics[angle=-90]{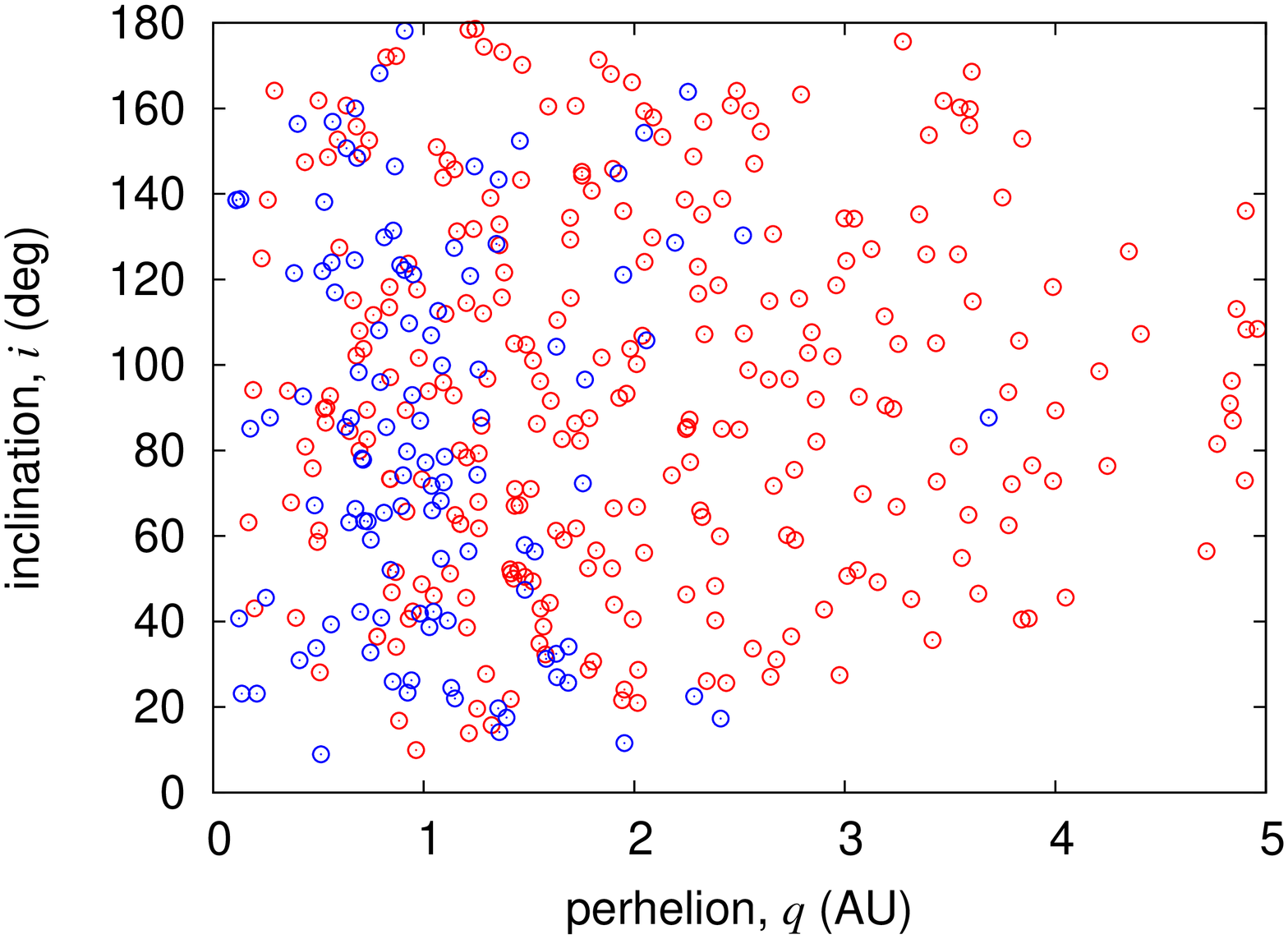}}
\resizebox{\hsize}{!}{\includegraphics[angle=-90]{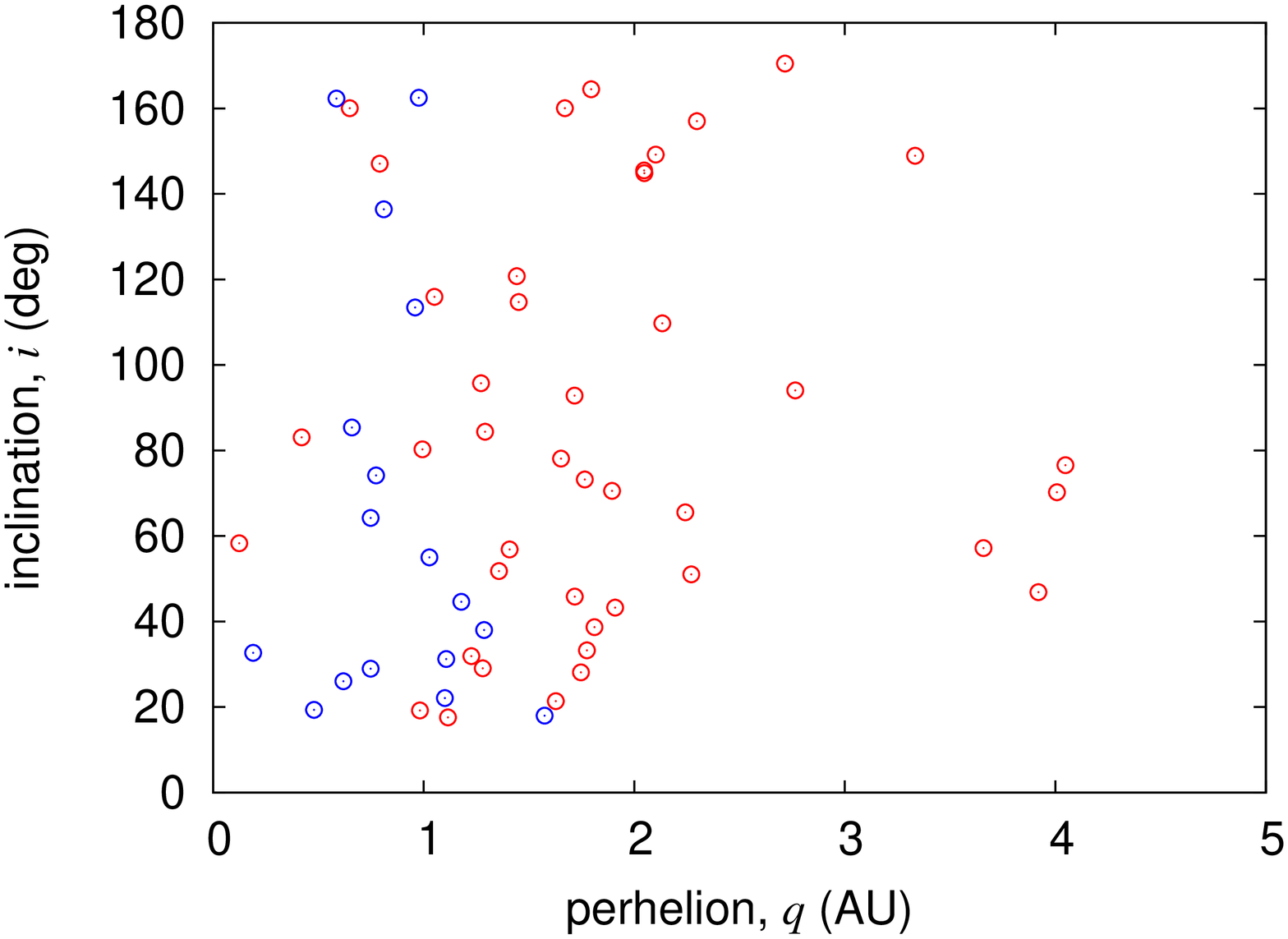}}
\caption{Perihelion versus inclination of comets from JPL Small-Body Search Engine. Top panel (plot only comets with $q<5$~AU, 92\% of
all JPL LPCs) is the plot of LPCs, and the bottom panel (plot only comets with $q<5$~AU, 100\% of all JPL HTCs) is for HTCs. Red circles
are comets discovered after 1950, blue circles are comets discovered before 1950. One can tell that, comets discovered before 1950 were
all concentrated at low perihelion distances.}
\label{fig:jpl_q_i}
\end{figure}

\begin{table} [ht]
%\begin{center}
\caption{A total of 60 HTCs from the JPL Small-Body Search Engine meet our HTC definition. There are 6 comets (marked with *)  
that were discovered before 1950, but could not be seen again between 1950 and 2013 due to their orbital periods are longer than
63 years.} 
\scriptsize
\vspace{4mm}
\begin{center}
\begin{tabular}{rrrrrrl}
\hline
$a$ & $e$ & $i$ & $q $ & $T_J$ & year & Name\\
(AU) & & ($^\circ$) & (AU) & &  \\
\hline
17.8 & 0.97 & 162.3 & 0.6 & -0.6 & 1758 & 1P/Halley \\
5.7 & 0.82 & 55.0 & 1.0 & +1.6 & 1790 & 8P/Tuttle \\
17.1 & 0.95 & 74.2 & 0.8 & +0.6 & 1812 & 12P/Pons-Brooks \\
16.9 & 0.93 & 44.6 & 1.2 & +1.2 & 1815 & 13P/Olbers \\
17.1 & 0.97 & 19.3 & 0.5 & +1.1 & 1847 & 23P/Brorsen-Metcalf \\
9.2 & 0.92 & 29.0 & 0.7 & +1.5 & 1818 & 27P/Crommelin \\
28.8 & 0.97 & 64.2 & 0.7 & +0.6 & 1788 & 35P/Herschel-Rigollet* \\
11.2 & 0.86 & 18.0 & 1.6 & +1.9 & 1867 & 38P/Stephan-Oterma \\
10.3 & 0.91 & 162.5 & 1.0 & -0.6 & 1866 & 55P/Tempel-Tuttle \\
3.0 & 0.96 & 58.3 & 0.1 & +1.9 & 1986 & 96P/Machholz 1 \\
26.1 & 0.96 & 113.5 & 1.0 & -0.3 & 1862 & 109P/Swift-Tuttle\\ 
17.7 & 0.96 & 85.4 & 0.7 & +0.4 & 1846 & 122P/de Vico \\
5.6 & 0.70 & 45.8 & 1.7 & +2.0 & 1983 & 126P/IRAS \\
7.7 & 0.84 & 95.7 & 1.3 & +0.5 & 1984 & 161P/Hartley-IRAS \\
24.3 & 0.95 & 31.2 & 1.1 & +1.3 & 1889 & 177P/Barnard \\
6.9 & 0.82 & 29.1 & 1.3 & +1.9 & 1994 & 262P/McNaught-Russell\\ 
32.8 & 0.98 & 136.4 & 0.8 & -0.6 & 1827 & 273P/Pons-Gambart* \\
23.6 & 0.95 & 56.0 & 1.2 & +1.0 & 1906 & C/1906V1(Thiele)* \\
27.6 & 0.99 & 32.7 & 0.2 & +0.6 & 1917 & C/1917F1(Mellish)* \\
7.9 & 0.86 & 22.1 & 1.1 & +1.8 & 1921 & C/1921H1(Dubiago) \\
32.7 & 0.98 & 26.0 & 0.6 & +1.0 & 1937 & C/1937D1(Wilk)* \\
19.4 & 0.93 & 38.0 & 1.3 & +1.4 & 1942 & C/1942EA(Vaisala)* \\
28.5 & 0.95 & 51.8 & 1.4 & +1.1 & 1984 & C/1984A1(Bradfield1) \\
18.9 & 0.98 & 83.1 & 0.4 & +0.4 & 1989 & C/1989A3(Bradfield) \\
13.8 & 0.93 & 19.2 & 1.0 & +1.5 & 1991 & C/1991L3(Levy) \\
12.1 & 0.82 & 109.7 & 2.1 & -0.2 & 1998 & C/1998G1(LINEAR) \\
23.0 & 0.92 & 28.1 & 1.7 & +1.6 & 1998 & C/1998Y1(LINEAR) \\
16.3 & 0.76 & 46.9 & 3.9 & +1.9 & 1998 & C/1999E1(Li) \\
28.1 & 0.86 & 76.6 & 4.0 & +0.7 & 1999 & C/1999G1(LINEAR)\\ 
17.0 & 0.92 & 120.8 & 1.4 & -0.4 & 1999 & C/1999K4(LINEAR) \\
18.9 & 0.90 & 70.6 & 1.9 & +0.8 & 1999 & C/1999S3(LINEAR) \\
17.2 & 0.87 & 157.0 & 2.3 & -1.4 & 2000 & C/2000D2(LINEAR) \\
14.0 & 0.81 & 170.5 & 2.7 & -1.5 & 2000 & C/2000G2(LINEAR) \\
13.3 & 0.93 & 80.2 & 1.0 & +0.6 & 2001 & C/2001OG108(LONEOS) \\
8.0 & 0.82 & 56.9 & 1.4 & +1.4 & 2001 & P/2001Q6(NEAT) \\
17.9 & 0.94 & 115.9 & 1.1 & -0.3 & 2001 & C/2001W2(BATTERS)\\ 
9.9 & 0.77 & 51.0 & 2.3 & +1.6 & 2001 & C/2002B1(LINEAR) \\
9.8 & 0.79 & 145.5 & 2.0 & -0.9 & 2002 & C/2002CE10(LINEAR) \\
17.5 & 0.84 & 94.1 & 2.8 & +0.2 & 2002 & C/2002K4(NEAT) \\
20.7 & 0.81 & 70.2 & 4.0 & +1.1 & 2003 & C/2003F1(LINEAR)\\ 
19.7 & 0.89 & 149.2 & 2.1 & -1.2 & 2003 & C/2003R1(LINEAR)\\ 
22.9 & 0.92 & 164.5 & 1.8 & -1.3 & 2003 & C/2003U1(LINEAR) \\
25.1 & 0.93 & 78.1 & 1.7 & +0.5 & 2003 & C/2003W1(LINEAR) \\
28.7 & 0.94 & 21.4 & 1.6 & +1.6 & 2005 & C/2005N5(Catalina) \\
23.7 & 0.86 & 148.9 & 3.3 & -1.6 & 2005 & C/2005O2(Christensen)\\ 
9.3 & 0.93 & 160.0 & 0.6 & -0.4 & 2005 & P/2005T4(SWAN) \\
7.8 & 0.84 & 31.9 & 1.2 & +1.8 & 2005 & P/2006HR30(SidingSpring)\\ 
5.6 & 0.70 & 160.0 & 1.7 & -0.5 & 2006 & P/2006R1(SidingSpring) \\
18.4 & 0.90 & 43.2 & 1.9 & +1.5 & 2008 & C/2008R3(LINEAR) \\
6.7 & 0.45 & 57.2 & 3.7 & +1.9 & 2010 & P/2010D2(WISE) \\
8.1 & 0.78 & 38.7 & 1.8 & +1.9 & 2010 & P/2010JC81(WISE)\\ 
8.2 & 0.90 & 147.1 & 0.8 & -0.3 & 2010 & C/2010L5(WISE) \\
19.6 & 0.93 & 114.7 & 1.5 & -0.3 & 2011 & C/2011J3(LINEAR)\\ 
11.0 & 0.80 & 65.5 & 2.2 & +1.2 & 2011 & C/2011L1(McNaught)\\ 
16.3 & 0.93 & 17.6 & 1.1 & +1.5 & 2011 & C/2011S2(Kowalski) \\
16.1 & 0.89 & 92.8 & 1.7 & +0.2 & 2012 & C/2012H2(McNaught)\\ 
8.5 & 0.85 & 84.4 & 1.3 & +0.7 & 2012 & P/2012NJ(LaSagra) \\
14.4 & 0.88 & 33.3 & 1.8 & +1.7 & 2012 & C/2012T6(Kowalski) \\
29.4 & 0.94 & 73.2 & 1.8 & +0.6 & 2012 & C/2012Y3(McNaught)\\ 
6.5 & 0.68 & 144.9 & 2.0 & -0.5 & 2013 & P/2013AL76(Catalina) \\
\hline
\end{tabular}
\end{center}
\label{tab:jplhtc}
\end{table}
One can imagine that, due to instrumental limitations, comets discovered even as late as the early 20th century needed to be
either very bright and/or very close to Earth, and hence were most likely observationally biased \citep{Everhart1967}. To 
support our claim, the perihelion distance versus inclination for LPCs (top) and HTCs (bottom) is shown in \fig{fig:jpl_q_i}. Blue
circles are comets discovered before 1950, and red circles are comets discovered after 1950. For both LPCs and HTCs the blue circles
were concentrated near low perihelia. Thus, it appears that comets discovered before 1950 had a stronger observational bias toward
low perihelia due to the instrumental limitations and limited sky coverage at the time. Comets discovered after 1950 are mostly
detected by surveys that have fewer biases  such as sweeping larger portions of the sky, and higher limiting magnitude. In addition,
the HTC sample contains more blue dots at low inclination than at high inclination, hinting at an inclination bias as well.
It has been well known that the HTCs show a predominantly prograde inclination distribution \citep{1994A&A...281..911F, Levison2001}.
It is possible that observational biases play a substantial role in the determination of the cumulative inclination distribution of
HTCs and this could be the reason why the median inclination value has thus far been substantially lower than the one expected from an
isotropic distribution \citep{Levison2001,Levison2006}. \cite{Levison2001} and \cite{Levison2006} tried to achieve observational
completeness by concentrating only on HTCs with $q< 1.3$~AU or 2.5~AU, following \cite{1988ApJ...328L..69D} and
\cite{1990ApJ...355..667Q}. However, it is not clear whether this prograde excess of HTCs is a result of observational biases or their
definition. Indeed, the relation $T_{\rm J} < 2$ implies $\sqrt{q}\cos i$ is approximately constant if $a\gg q$, which holds for most
HTCs. Thus using an HTC catalogue with a maximum $q$ of 1.3~AU may result in a lower median inclination than when imposing a maximum
$q$ of 2~AU. This phenomenon is clearly visible in Table~\ref{tab:jplhtcnumberq} for the observational data, which lists the number of
discovered HTCs in 50-year intervals, together with their median perihelion distance and median inclination. Dates closer to the
current epoch detect HTCs with higher perihelion distance because of their increased limiting magnitude and sky coverage, but the
median inclination also increases and approaches an isotropic distribution. However, data from our numerical simulations,
discussed below, contain no such trend and instead have a median inclination close to $90^\circ$ for all values of $q$, supporting the
idea that the prograde excess is caused by early observational biases.\\
\begin{table} []
\caption{The number of HTCs discovered every 50 years starting from 1850 to 2013. 29 HTCs were found between 2000
to 2013.} 
%\scriptsize
\vspace{4mm}
\begin{center}
\begin{tabular}{rccc}
\hline
 year & number & median $q$ (AU) & observed median $i$ (deg)\\
\hline
1850-1900 & 4 & 1.0 & 72\\
1900-1950 & 5 & 1.1 & 33\\
1950-2000 & 13 & 1.4 & 58\\
$>$2000 & 29 & 1.8 & 84\\
\hline
\end{tabular}
\end{center}
\label{tab:jplhtcnumberq}
\end{table}
At the same time we cannot rule out that the definition of HTCs is solely responsible for the prograde excess. Indeed, none of
the previous studies made a cut to the dataset based on the epoch of observations, potentially still leading to a biased sample. Here
we make use of JPL catalogue from \fig{fig:jpl_q_i} to verify this point. Apparently, if one only keeps all HTCs with $q<1.3$~AU by arguing
that this yields an observationally complete sample for the entire dataset, then many new discoveries due to advanced technology after
1950 will be discarded. Not knowing the observational conditions for comets discovered before 1950, attempting to de-bias the
old discoveries would be difficult. Fortunately, we have many new discoveries in the last 20 years see Table~\ref{tab:jplhtcnumberq})
and therefore we decided that the easiest way to de-bias the old discoveries is to remove comets discovered earlier than 1950 that did
not pass perihelion between 1950 and 2013.\\

\begin{figure}[ht]
\begin{center}
\resizebox{\hsize}{!}{\includegraphics[angle=-90]{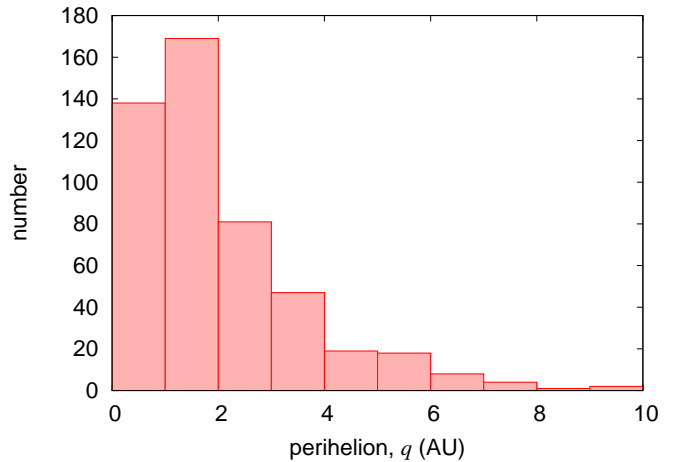}}
\end{center}
\caption{The histogram of perihelion distance $q$ of comets with $T_J<2$ from JPL catalogue. It shows a sharp drop of number of comets
with $q>2$ AU.}
\label{fig:Qhistogram}
\end{figure}
The last issue that warrants discussion is the maximum imposed perihelion distance of the observed sample. \cite{1988ApJ...328L..69D}
and \cite{Levison2001} imposed different threshold values of the maximum perihelion distances below which the comet sample was thought
to be observationally complete. In other words, it was assumed that below a maximum perihelion distance $q_{\rm m}$ there existed a
100\% detection efficiency, up to a certain total limiting magnitude, for all comets. For this we chose to keep the maximum perihelion
distance a free parameter below 2~AU. Our choice for the upper limit is motivated by the distribution of perihelion distance of comets
with $T_j<2$ in JPL catalogue, which shows a sharp decrease beyond 2~AU (see \fig{fig:Qhistogram}).\\
We ended up with a total of 158 LPCs and 40 HTCs. We now have a fairly unbiased sample of LPCs and HTCs from the JPL comet database for
comparison between theory and observations. Thus we move on to describe our numerical methods in the next section.

\section {Numerical simulation and Initial Conditions}
We determined whether or not the Oort cloud is a feasible mechanism to generate the HTCs by running numerical simulations of the
evolution of Oort cloud comets under the gravitational perturbations of the Sun, giant planets, Galactic tide and passing field stars
and comparing the ensuing orbital element distributions against those of the observed sample.\\
We generated a fictitious Oort cloud by using the method described in \cite{2009ApJ...695..268K}  from the Oort cloud formation 
simulations presented in \cite{2013Icar..225...40B}. We divided each Oort cloud in six equal sections of semi-major axis and computed
the cumulative distributions of the argument of perihelion ($\omega)$, longitude of the ascending node ($\Omega$), specific orbital
angular momentum ($L)$ and the $z$-component of the specific orbital angular momentum ($L_z$) for each section. The cumulative
distribution of the semi-major axis was computed for the whole Oort cloud. We generated a comet by picking a uniform random number
between 0 and 1 and selecting the corresponding semi-major axis from the semi-major axis distribution. The value of the semi-major axis
identified the section of the cloud the comet would reside in. We then picked another four uniform random numbers to determine the
values of $\omega$, $\Omega$, $L$ and $L_z$. The eccentricity was then computed from $L$ and the inclination from $L_z$. The mean
anomaly of the comet was picked at random uniformly between 0 and 360$^\circ$. We repeated this process 32768 times for each simulation
and generated 40 separate simulations, resulting in a total of 1.3 million test particles.\\
The motion of the comets was subsequently integrated using the SWIFT RMVS3 integrator \citep{Levison1994} without the giant planets
present, but included the perturbations from the Galactic tide and the passing stars. The perturbations from the Galactic tide and
stars were modelled after \cite{Levison2001} for the tides and \cite{2008CeMDA.102..111R} for the stars. The comets were removed from
the simulation once they got closer than 38~AU from the Sun or farther than 1~pc from the Sun. These simulations were run for 4~Gyr
with a time step of 50~yr. Different stellar encounter files were used for each simulation.\\
The second part of our simulations consisted of re-integrating only those comets that were removed from the first set of simulations
because they came closer than 38~AU from the Sun. We also integrated a single clone of these objects to increase the total particle
count per simulation. On average only 18\% of all Oort cloud objects penetrated the planetary region; the rest remained in the Oort
cloud or were removed by other means. The clone had the same original elements apart from a random mean anomaly. This typically
resulted in 12~000 test particles per simulation and a total of 40 simulations. The difference with the first set of simulations is
that now we included the giant planets. This procedure of running the simulations in two parts increases the number of comets we can
integrate through the realm of the giant planets.\\
We ran these second set of simulations with SCATR \citep{Kaib2011} which is a Symplectically-Corrected Adaptive Timestepping Routine.
It is based on SWIFT's RMVS3 but it has a speed advantage over SWIFT's RMVS3 or MERCURY \citep{Chambers1999} for objects far away from
both the Sun and the planets where the time step is increased. We set the boundary between the regions with short and long time step at
300~AU from the Sun \citep{Kaib2011}. Closer than this distance the computations are performed in the heliocentric frame, like
SWIFT's RMVS3, with a time step of 0.4~yr. Farther than 300~AU, the calculations are performed in the barycentric frame and we
increased the time step to 50~yr. The error in the energy and angular momentum that is incurred every time an object crosses the
boundary at 300~AU is significantly reduced through the use of symplectic correctors \citep{Wisdom1996}. For the parameters we
consider, the cumulative error in energy and angular momentum incurred over the age of the Solar System is of the same order or smaller
than that of SWIFT's RMVS3. The same Galactic and stellar parameters as in the first simulation were used. Comets were removed once
they were further than 1~pc from the Sun, or collided with the Sun or a planet.\\
To determine how the comets fade we modified SCATR to keep track of the number of perihelion passages, $n_q$, that each comet made.
\cite{Levison2001} suggest that comets fade strongest when their perihelion distance $q<2.5$~AU, so we only counted the perihelion
passages when the perihelion was closer than 2.5~AU. The number of passages together with time, osculating barycentric semi-major axis,
eccentricity, inclination, perihelion distance and mean anomaly were stored in a separate file with an output interval of 100~yr but
only when the comet was closer than 300~AU from the Sun. The format of the output made for easy comparison with the JPL catalogue. \\
The sublimation of the cometary material due to the solar radiation will make the comets become fainter and fainter after each
perihelion passage. Depending on the nature of the cometary material, comets will eventually become too faint after a high number
of perihelion passages and turn into dormant or extinct comets that easily escape detection. This is the so called fading problem
first suggested by \cite{1950BAN....11...91O} to explain the deficiency of observed comets. A more extended discussion of different
fading functions can be found in \cite{Wiegert1999}. In this work, we applied a post-processing simple power-law fading of the form
\begin{equation}
\Phi_{m}=m^{-k},
\label{equ:fading}
\end{equation}
where $\Phi_m$ is the remaining visibility function introduced by \cite{Wiegert1999}. A comet that has its $m^{\rm th}$
perihelion passage with $q<2.5$~AU will have its remaining visibility be $\Phi_m$, where $k$ is a positive constant, and $\Phi_1=1$ for
the first out-going perihelion passage of a comet. Without fading, every comet in our simulation has the same weighting ($\Phi_m=1,\,\,
m=1,2,3 \ldots, n_{\rm p}$) in constructing the cumulative distribution of semi-major axis or inclination of active comets. When we
applied the fading effect to comets, the remaining visibility $\Phi_m$ is considered as a weighting factor. The larger the number of
perihelion passages, the less each comet contributes to the cumulative inclination or semi-major axis distribution of the active
comets.\\
For example, in constructing the inclination distribution with fading, we first sort the comets by inclination in increasing order and
make the cumulative fraction based on their remaining visibilities. In this way we can construct the cumulative distribution of
orbital elements with fading by changing the power law index $k$ in the remaining visibility. \cite{Whipple1962} and \cite{Wiegert1999}
found $k \sim 0.6$ has the best match between simulation and observation and thus we expect to obtain a similar value here.  
Our database has a baseline of 63 years, which is shorter than the period of some HTCs. Therefore, before comparing our simulation to
the observation, the cumulative distributions of inclination and semi-major axis from simulation have to be weighted by the ratio of
the period of comets and the observed window. We crudely corrected for the bias with a weighting factor to the remaining visibility
by, 
\begin{equation}
\Phi_{m}'=m^{-k}\Bigl(\frac{63\,{\rm yr}}{a^{3/2}}\Bigr); \;\; a>15.8 \,\rm{AU.}
\label{equ:weighting}
\end{equation}
For every discovered comet with high semi-major axis ($a>15.8$ AU), there are actually $a^{3/2}/63$ more comets of similar semi-major
axis comets. We ran all of our simulations on the CP computer cluster at ASIAA. The total time taken per simulation was about one
month.

\section{Simulation Results}
In this section we present the results of our numerical simulations and the comparison with observational data. Our simulations yielded
7140 individual LPCs  and 1159 individual HTCs that pass the same criteria, with each category containing a few million entries.

\subsection{Oort cloud model and fading}
We first need to establish whether our OC model is compatible with the observed OOCs and LPCs. We followed the analysis in
\cite{Wiegert1999} to determine whether or not our simulations could simultaneously match different parts of the observed LPC
population from JPL catalogue. \cite{Wiegert1999} introduce three quantities that measure the relative contributions to the flux of
comet passing perihelion. These are: $\Psi_1$ is the fraction of all comets with $a>10$~kAU, $\Psi_2$ is the
fraction of all comets that have $34.2<a<69.8$~AU, i.e., the tail end of the LPC distribution, and $\Psi_3$ is the fraction of all comets
that are prograde. In addition to the $\Psi$ values, we also follow \cite{Wiegert1999} and define the simulated to observed
ratio $X_i=\Psi_i^\prime / \Psi_i$, where $\Psi_i^\prime$ is the simulated value after applying fading. A good model for the
Oort cloud should yield $X_i=1$ simultaneously at certain fading index $k$.\\
In order to compare the flux of comets in our simulations with those of \cite{Wiegert1999} we need to impose a maximum perihelion
distance for the LPCs, as opposed to the subsequent HTC analysis, where we consider them to always be detected. For our purposes we
decided to use $q_{\rm m} = 2.0$~AU which equals the median value in the current LPC sample.  From the JPL catalogue for comets with
$q<2.0$~AU which are discovered or observed after 1950 we have
\begin{itemize}
\item $\Psi_{1}=0.044 \pm 0.020$,
\item $\Psi_{2}=0.152 \pm 0.043$,
\item $\Psi_{3}=0.519 \pm 0.100$,
\end{itemize}
where the errors are assumed to be Poisson statistics. We did not include hyperbolic comets in the determination of $\Psi_1$.
The $\Psi$ values should be compared to those in \cite{Wiegert1999}, who reported $\Psi_1 = 0.380 \pm 0.043$, $\Psi_2 = 0.066 \pm
0.015$ and $\Psi_3 = 0.505 \pm 0.051$. The largest discrepancy occurs in $\Psi_1$, with our value approximately an order of magnitude
lower than theirs. We attribute this to the fact that \cite{Wiegert1999} used the Catalogue of Cometary Orbits of 1993
\citep{Marsden1993} rather than JPL's, and they calculated $\Psi_1$ from the original semi-major axis of the comets. In addition,
\cite{Wiegert1999} did not apply a cut in $q$ when calculating $\Psi_1$ and used all the LPCs in the catalogue, which could also
increase the value of $\Psi_1$.\\
Here we decided to use the JPL catalogue, which publishes the semi-major axis of comets when they are still in the realm of the giant
planets. Doing this skews the semi-major axis distribution with the result that the contribution from the Oort spike is substantially
decreased. This reduction in the Oort spike is the result of planetary perturbations which cause a change in binding energy
that is typically greater than the original binding energy \citep{1981A&A....96...26F}. Thus the use of the JPL catalogue and evaluating
the orbits in the realm of the giant planets may cause errors in classifying a comet as LPC or OCC. Thus we need to determine whether
or not our low value of $\Psi_1$ is somehow consistent with that of \cite{Wiegert1999} by correcting for the different methods in
publishing the semi-major axes.\\
Even though the initial conditions of \cite{Wiegert1999} were very different from ours, we can use the results of their numerical
simulations to estimate our value of $\Psi_1$ when planetary perturbations are removed. Without cometary fading \cite{Wiegert1999} find
$X_1 = 0.075 \pm 0.011$, and because $\Psi_1^\prime = \Psi_1 X_1$ our corrected value is $\Psi_1 \sim 0.508 \pm 0.185$. This is
consistent with \cite{Wiegert1999} within the error bars.\\
\begin{figure}[ht]
\begin{center}
\resizebox{\hsize}{!}{\includegraphics[angle=-90]{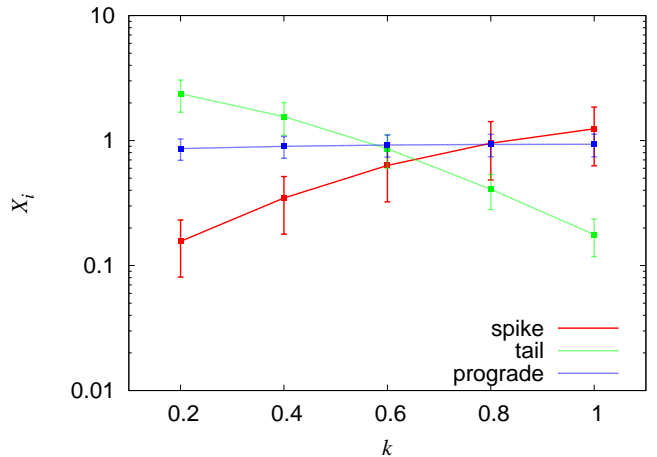}}
\end{center}
\caption{The values of $X_i$ as a function of fading index $k$. The three curves cross each other near the region
$ 0.6<k<0.8$, indicating that after applying fading to the simulations, the best match should fall in this region.}
\label{fig:Xvalue}
\end{figure}
\cite{Wiegert1999} report that numerical simulations without fading yield values of $X_i$ that are very different from unity, with
$X_1$ being too low and $X_2$ being too high. For this reason they introduce several fading laws and compute the $X_i$
values for various fading parameters. They argue that introducing a fading law should result in all $X_i$ values being near unity, with
some laws better than others. They obtain the best result from the power law fading equation (\ref{equ:fading}) above with $k \sim
0.6$. We applied the same fading law to our simulations to test our initial conditions and numerical methods. In \fig{fig:Xvalue}
we depict the $X$ values as a function of fading index $k$. The red line plots $X_1$, the green line is for $X_2$ and the blue line
depicts $X_3$. One can tell from the plot that after applying the fading, our simulations are a good match to the observed distribution
when the fading index is in the range $0.6<k<0.8$, with a slight preference towards to 0.6 than to 0.7. This is close to the same
value found by \cite{Whipple1962} and \cite{Wiegert1999} and suggests that our Oort cloud model can be made consistent with
observations if the comets fade via this law.\\
The result depicted in \fig{fig:Xvalue} begs the following question: can the same fading law be used to reproduce the observed orbital
distribution of HTCs? We attempt to answer this in the next subsection.

\subsection{The Halley-type comets}
In order to find out how well our simulations match with  the observed HTCs, we computed the cumulative inclination and semi-major
axis distributions of the observed and simulated comets for a range of maximum perihelia, $q_{\rm m}$, and fading parameter, $k$. Once
these distributions were generated we performed a Kolmogorov-Smirnov (K-S) test \citep{1992nrca.book.....P}, which searches for the
maximum absolute deviation $D_{\rm max}$ between the observed and simulated cumulative distributions. The K-S test assumes
that the entries in the distributions are statistically independent. The probability of a match, $P_D$, as a function of $D_{\rm
max}$, can be calculated to claim whether these two populations were from the same parent distribution.\\ 
However, in our simulations, a single comet would be included many times in the final distribution as long as the comet met our
criteria for being an HTC during each of its perihelion passages. Including its dynamical evolution in this manner would cause
many of the entries in the final distribution to become statistically dependent and the K-S test would be inapplicable. In the
meantime, there are only  between 28 and 40 HTCs in the JPL catalogue that passed our criteria, depending on $q_{\rm m}$. Thus we may
suffer from small number statistics. These limitations prevent us from using the analytical formula described in
\cite{1992nrca.book.....P} to calculate the K-S probability.\\
We solved this issue by applying a Monte Carlo method to perform the K-S test as described in \cite{Levison2006}. Once we have the
inclination and semi-major axis distributions from the simulation, we then randomly selected 10\,000 fictitious samples from the
simulation. Each sample has the same number of data points as HTCs from JPL catalogue. The Monte-Carlo K-S probability is then the
fraction of cases that have their $D$ values between the fictitious samples and real HTC samples larger than the $D_{\rm max}$ found
from the real HTC samples and cumulative distributions from simulation. \\
However, before making the fictitious datasets, we need to find the empirical probability density functions of inclination and
semi-major axis from which we then sample the fictitious HTCs. Here we generated these from a normalized histogram. One crucial
point in making the histogram is that we weighed each entry by its remaining visibility (equation \ref{equ:fading}) and its
orbital period (equation \ref{equ:weighting}). The fictitious comets were then sampled from the distributions with the von Neumann 
rejection technique \citep{vonNeumann1951}. This sampling method relies on generating two uniform random numbers on a grid. An entry is
accepted when both numbers fall under the probability density curve.\\
\begin{figure}[ht]
\resizebox{\hsize}{!}{\includegraphics[angle=-90]{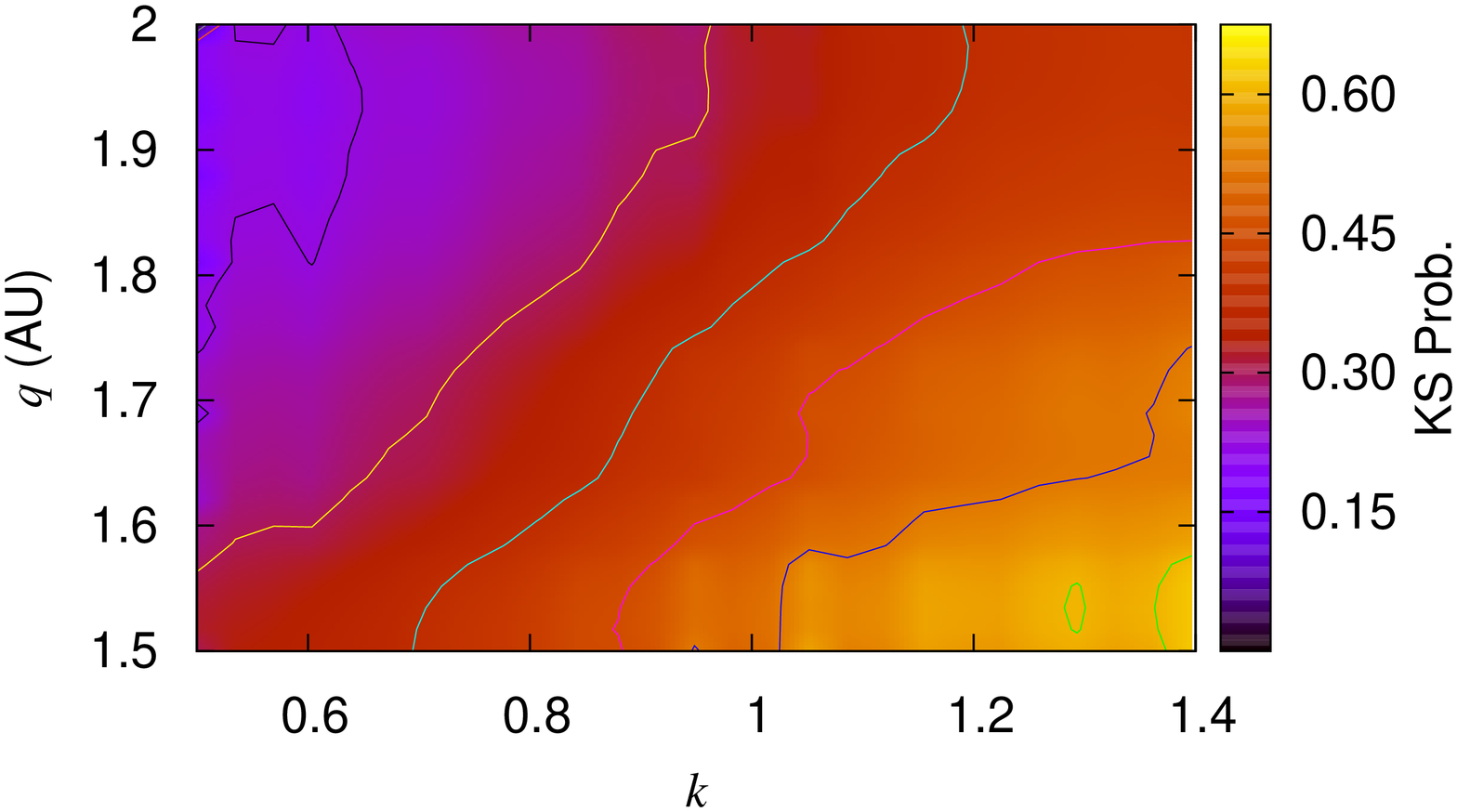}}
\resizebox{\hsize}{!}{\includegraphics[angle=-90]{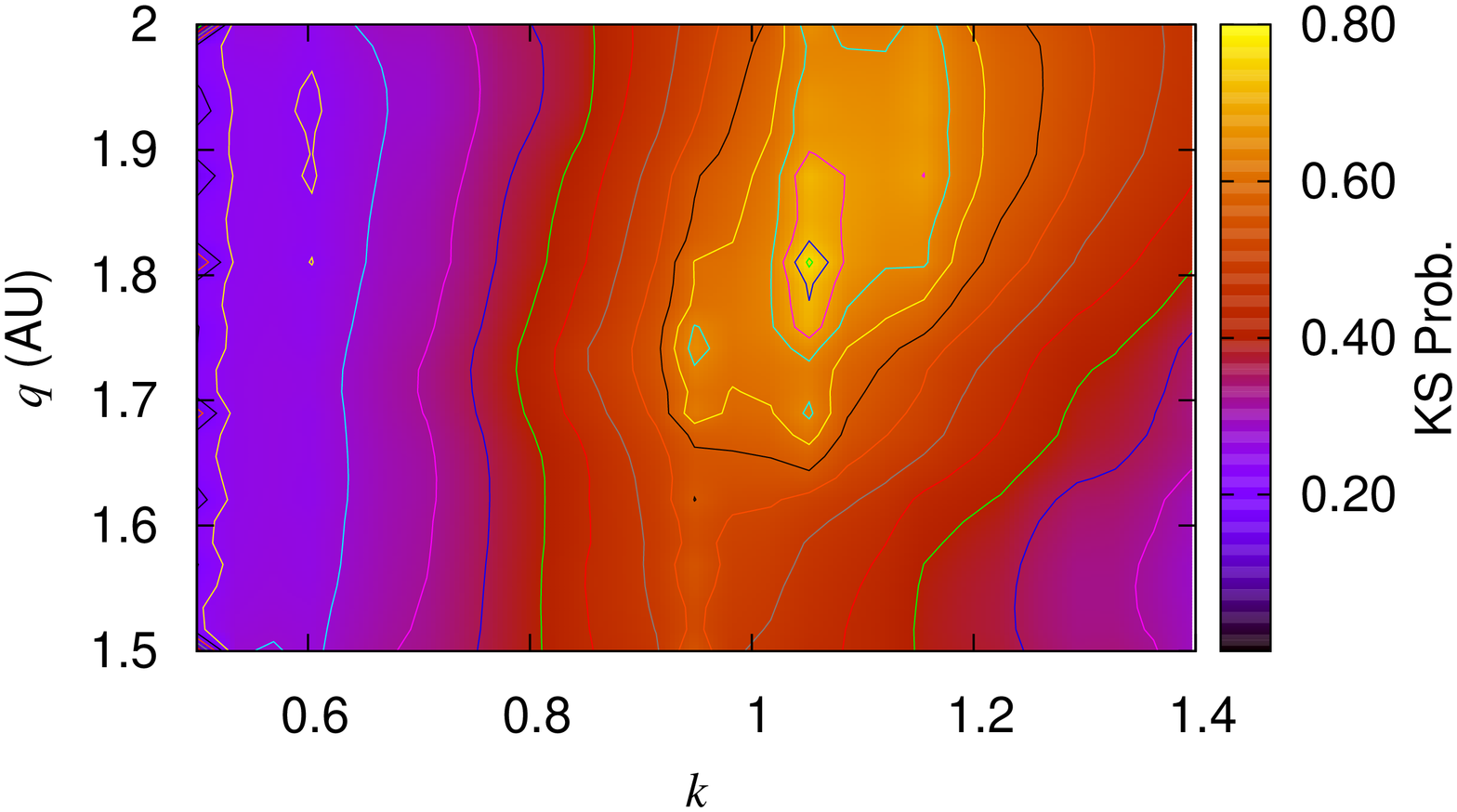}}
\caption{Contour plots of the K-S probability as a function of the fading parameter and maximum perihelion distance. The top panel
shows the fit for the inclination distribution. The bottom panel pertains to the semi-major axis. The parameters that best fit both
distributions are $k=1.05$ and $q = 1.77$~AU.}
\label{fig:result_htc_ia_contour}
\end{figure}
\begin{figure}[ht]
\resizebox{\hsize}{!}{\includegraphics[angle=-90]{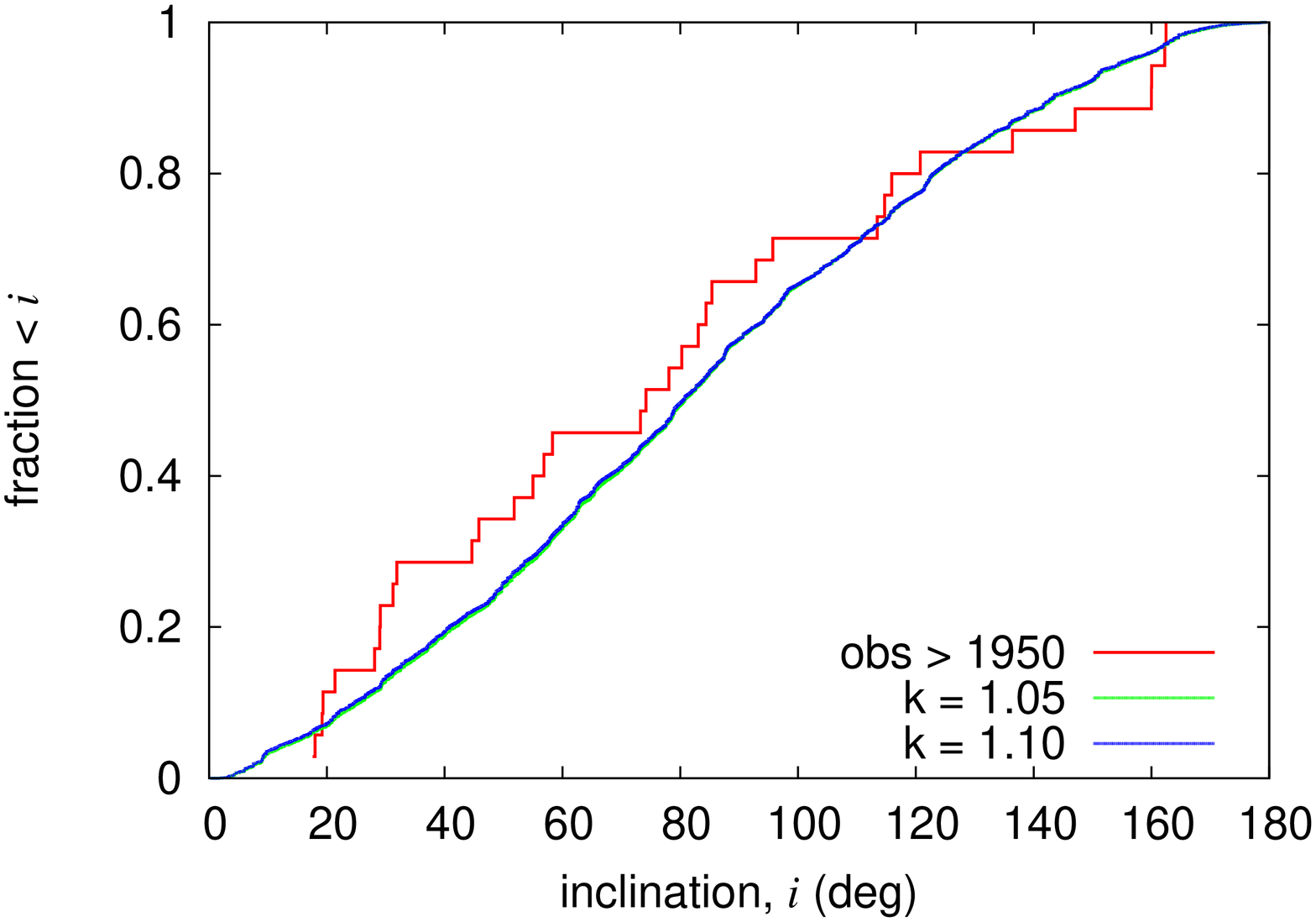}}
\resizebox{\hsize}{!}{\includegraphics[angle=-90]{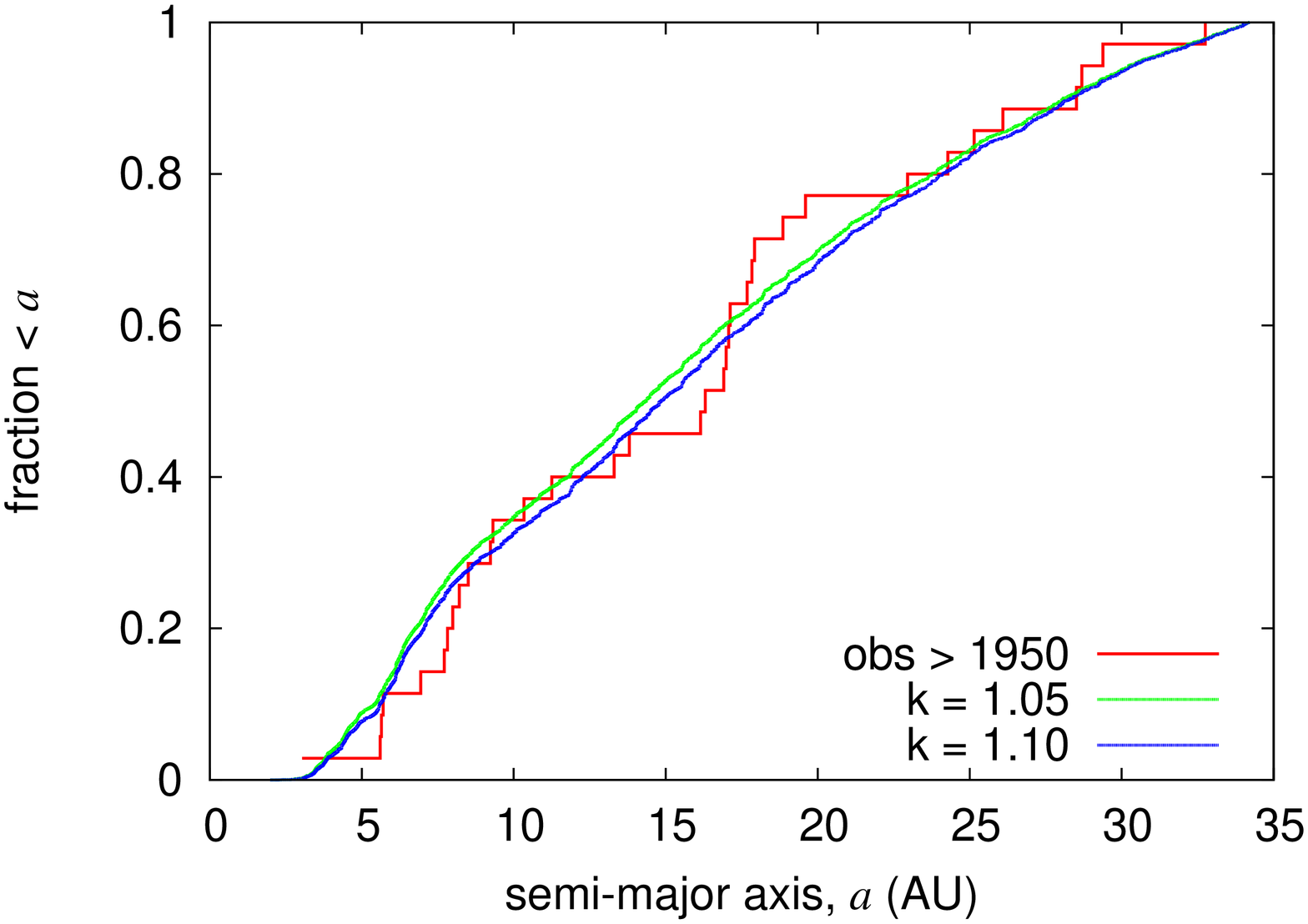}}
\caption{The inclination and semi-major axis distributions from observation and simulation for HTCs that gave the best match from
Fig.~\ref{fig:result_htc_ia_contour} above. }
\label{fig:result_htc_ia}
\end{figure}
The results of applying the fading to the simulated data in the HTC region are shown in \fig{fig:result_htc_ia_contour}. Here we
plot contours of the K-S probability as a function of $k$ and $q_{\rm m}$ for the inclination distribution (top) and semi-major axis
distribution (bottom). The highest K-S probability for the inclination occurs at low $q_{\rm m}\sim1.5$~AU and relatively high
$k\sim1.4$, while for the semi-major axis distribution the maximum K-S probability occurs when $q_{\rm m} \sim 1.8$~AU and $k\sim 1$.
Combining these results we find that the combination of $q_{\rm m}$ and $k$ that best fits both the semi-major axis and inclination
distribution is $q=1.77$~AU, $k=1.05$ with K-S probabilities of 84.3\% for the semi-major axis and 31\% for the inclination.
We show the resulting fits to the data in \fig{fig:result_htc_ia} below. The top panel shows the inclination distribution, the bottom
panel depicts the semi-major axis distribution. The median simulated inclination is 80.4$^\circ$ (observed 74.2$^\circ$) and the median
simulated semi-major axis is 14.3~AU (observed 16.3~AU). \cite{Levison2001} apply a different fading law than the simple one we
proposed here but they conclude that their model may work when $k \sim 0.8$. This is of the same ballpark as ours, but they state that
their value is very uncertain. \\
Summarizing, it appears that an Oort cloud source combined with a simple fading law can match the observed HTC inclination and
semi-major axis distribution rather well  when imposing a maximum perihelion distance close to 1.8~AU and taking the fading
parameter to be $\sim 1$. The imposed upper value of the perihelion distance is below the maximum of 2~AU beyond which the catalogue
shows a steep decrease in the number of comets, and is close to the median value obtained from observations in the last 13 years. This
suggests that the current catalogue of HTCs with $q<1.8$~AU is nearing completion.\\
Unfortunately the optimum value of $k$ is inconsistent with that of the LPCs found in the previous subsection. It appears that the
HTCs fade quicker than the LPCs. It is now well recognized that cometary nuclei develop non-volatile, lag-deposit crusts that reduce
the fraction of the nucleus surface available for sublimation \citep{Brin1979, Fanale1984}. For most Jupiter-family comets (JFCs), the
'active fraction', that is the fraction of the nucleus surface area that must be active to explain the comet's water production rate,
is typically only a few per cent, or even a fraction of a per cent \citep{1999A&A...352..327F}. Most likely for HTCs the active surface
fraction is similar to that of JFCs. Activity could be re-ignited if the comet undergoes a substantial decrease in perihelion
\citep{2008CeMDA.102..111R}. The age of the comet and the surface being non-pristine could account for the fading index being
higher than for new comets that first enter the inner Solar System.

\subsection{Expected HTC population}
We may use the results of the numerical simulations above to constrain the expected number of HTCs larger than a given size using the
method described in \cite{1997Icar..127...13L} and \cite{2013Icar..225...40B}. These works focused on the active lifetimes of
Jupiter-family comets (JFCs). The method consists of finding the best match of simulated comets to the observed semi-major axis
and inclination distributions as a function of active lifetime $\tau_{\rm{vHTC}}$. In other words one computes the K-S
probability for matching the simulated inclination or semi-major axis distribution to the observed one for each value of the
active lifetime. The maximum K-S probability corresponds to the best estimate of $\tau_{\rm{vHTC}}$. However, both of the above
studies assumed the comets retained their original activity for a time $\tau_{\rm vHTC}$ and would fade completely immediately
afterwards. Our matching procedure here must rely on a different approach because we fade the comets with time. Thus the choice of
$\tau_{\rm vHTC}$ should depend on the fading index $k$ and the number of perihelion passages after which we consider the comet to have
faded completely. There is also a $q$ dependence but in what follows we only consider HTCs with $q<1.8$~AU.\\
The formula relating the number of active HTCs ($N_{\rm{vHTC}}$) to the number of Oort cloud comets ($N_{\rm{OC}}$) is
\begin{equation}
 N_{\rm vHTC} = N_{\rm OC}\tau_{\rm vHTC}|r_{\rm OC}|f_{\rm vHTC},
\label{formula}
\end{equation}
where $r_{\rm{OC}}$ is the fractional decay rate of the Oort cloud at the current time, $f_{\rm{vHTC}}$ is the fraction of the
comets escaping from the Oort cloud that penetrate into the visible HTC region with $q<1.8$~AU and $N_{\rm vHTC}$ and $N_{\rm
OC}$ are the number of visible HTCs and Oort cloud comets. We evaluate these three quantities below.\\
Defining by $f_{\rm{OC}}(t)$ the fraction of Oort cloud objects surviving in the Oort cloud at time $t$, the fractional decay rate of
the Oort cloud population is defined as $r_{\rm{OC}}(t)=(df_{\rm{OC}}(t)/dt)/f_{\rm{OC}}(t)$. We measured $r_{\rm{OC}}$ from the
simulations in Brasser \& Morbidelli (2013) over the last 3~Gyr and obtained $\langle r_{\rm{OC}} \rangle =-1.48 \times 10^{-10}$ per
year.\\
The fraction of Oort cloud objects that became HTCs with $q<1.8$~AU, $f_{\rm{vHTC}}$, is tied to the active lifetime, $\tau_{\rm
vHTC}$. When the HTCs enter the inner solar system for the first time with $q<1.8$~AU many of them will have $n_q \gg 1$ and thus some
of these will already have faded and will not be detected. If we were to assume that the average time any HTC spends with $q<1.8$~AU as
an {\it active} object at each output entry equals the entire 100~yr between outputs we obtain $\tau_{\rm{vHTC}} = 310$~kyr, which is
longer than the median dynamical lifetime of 112~kyr (for comparison, \cite{Levison2006} reported a median dynamical lifetime of
69~kyr and \cite{2002MNRAS.333..835N} gave 100~kyr). Thus a different approach is needed.\\
\cite{2013Icar..225...40B} report an active JFC lifetime of $\tau_{\rm vJFC} =12$~kyr. This corresponds to roughly 2\,000 revolutions
but only $\sim$ 400 of these are spent with $q<2.5$~AU. Imposing the latter as an upper limit
on $n_{\rm p}$ for the HTCs results in a remaining visibility of the order of 0.25\% when $k \sim 1$, which implies the comet has  faded
substantially: the change in absolute magnitude is $\Delta H = \frac{5}{2}k\log n_{\rm p}=6.5$. We take this as a good upper limit to
consider the comet to have faded for the following reasons.\\
\cite{2011MNRAS.416..767S} showed that an LPC with total absolute magnitude $H_{\rm T} = 6.5$ has a
nuclear magnitude of $H_{\rm N} = 17.3$, corresponding to a diameter of 2.3~km assuming an albedo of 4\%. A JFC with the same size has
a typical total absolute magnitude $H_T=9.3$ \citep{1999A&A...352..327F, 2013Icar..225...40B}. Thus, on average, one may consider the
JFCs to have faded by 2.8 magnitudes compared to LPCs of the same size. Assuming that for the HTCs follow the same trend, our earlier
limit of 6.5 magnitudes for complete fading suggested above seems reasonable.\\
Assuming once again that the comet spends the full 100~yr in the region of interest between outputs, and also assuming that a similar
amount of time is spent with $q>2.5$~AU as for the JFCs \citep{2013Icar..225...40B}, we obtain $\tau_{\rm{vHTC}} = 23.2$~kyr.\\
We then compute $f_{\rm{vHTC}}$ by identifying all unique HTCs in our simulations that have $n_q \leq 2000$ when they first reach
$q<1.8$~AU and dividing by the total number of comets. This results in $f_{\rm{vHTC}} = 4.2 \times 10^{-4}$. Our value is
approximately a factor 1.5 lower than that found by \cite{Levison2001}.\\ 
The last ingredient of the formula consists of the number of comets in the Oort cloud. \cite{2013Icar..225...40B} report there are
$7.6 \times 10^{10}$ comets in the Oort cloud with diameter $D>2.3$~km. Combining these together results in $N_{\rm{vHTC}} \sim 100$
with $D>2.3$~km and $q<1.8$~AU. This estimate is quite uncertain, with the largest uncertainty being in $\tau_{\rm vHTC}$, which can
easily vary by a factor of a few. Incidentally the number of active HTCs is comparable to the number of active JFCs of the same size
reported in \cite{1997Icar..127...13L} and \cite{2013Icar..225...40B}. The dormant to active HTC population is approximately 4, given
by dividing the dynamical lifetime by the active lifetime.

\section{Discussion}
In the previous section we demonstrated the ability to reproduce the currently observed inclination and semi-major axis distribution
of the HTCs from an Oort cloud source using a simple cometary fading law. The result is simple, works very well and thus begs further
discussion.\\
We want to emphasize here that we have been careful in minimizing possible sources of error and contamination. First, we have
justified our rationale for using the JPL catalogue and comparing our simulations to the observational data. The initial conditions
from our simulations come from earlier Oort cloud formation simulations from \cite{2013Icar..225...40B} using a robust method for
obtaining a highly populated Oort cloud without needing to resort to perform many Oort cloud formation simulations 
\citep{2009ApJ...695..268K}. We have used a computer code that has been tested extensively and used the full models for the Galactic
tides and passing stars \citep{Levison2001, 2008CeMDA.102..111R}. Thus we see no obvious substantial difference in the numerical
methodology with earlier works that attempted to reproduce the orbit distribution of HTCs.\\
There are two issues where we differ from earlier attempts by \cite{Levison2001} and \cite{Levison2006}. First, we included more
HTCs in our observational data. Both \cite{Levison2001} and \cite{Levison2006} used the observed HTCs with $q<1.3$~AU, arguing that
HTCs with larger perihelion distance were observationally incomplete. Their sample remained small, however, containing just $\sim$20
comets. However, they also included comets discovered before 1950 which we have discarded if they did not pass perihelion after
1950 for reasons discussed in Section~2. Second, the last decade has seen a vast increase in the number of observed comets, including
new HTCs. Compared to \cite{Levison2001} and \cite{Levison2006}, we have different inclination and semi-major axis distributions
because of the cut in the epoch of observations, more new observations, and higher maximum perihelion distance. In other words, in
our work we were comparing our simulated comets to a different inclination and semi-major axis distributions than \cite{Levison2001}
and \cite{Levison2006} because of the increased number of comets in the last decade. Specifically, the median inclination in the sample
of \cite{Levison2006} is 55$^\circ$ and in our best-fit sample it is 75$^\circ$, inching towards a more isotropic distribution that
would be expected from the models. Similar arguments apply to the work of \cite{2002MNRAS.333..835N}, with the added emphasis that we
had greatly different initial conditions, in particular the initial perihelion distribution.\\
A second point that warrants discussion is our method of calculating the fading of the comets. We based the fading on the methods of
\cite{Wiegert1999} because of their surprising good result for power-law fading when $k\sim 0.6$. Unfortunately \cite{Wiegert1999} did
not describe how they applied their fading in great detail, something we did here. It is possible that our methods differ from those of
\cite{Wiegert1999} but the fact that we obtain similar results suggests otherwise.

\section{Summary and conclusions}
In this work we have used numerical simulations to determine whether the distribution of the HTCs can be reproduced using the Oort
cloud as their source distribution and when considering cometary fading. We generated a large number Oort cloud comets that
were evolved numerically under the gravitational influence of the Sun, giant planets, Galactic tide and passing stars using
well-established methods. Guided by the work of \cite{Wiegert1999} we post-processed the data with a power law fading mechanism that
was applied to all comets that passed perihelion closer than 2.5~AU. We compared the numerical output to the JPL catalogue,
keeping only those comets discovered and observed after 1950 to limit observational biases.\\
We find that the Oort cloud with cometary fading is the most likely source of the HTCs. Both of their semi-major axis and
inclination distributions can be well reproduced when the fading index $k\sim1$ and when one considers only those comets for
which $q \lesssim 1.8$~AU. The HTC fading index is somewhat higher than the value of 0.6 we find for LPCs. This increase in fading
index could be the result of physical processes on the cometary surface. With these results we estimate there are of the order of 100
active HTCs with diameters $D>2.3$~km and $q<1.8$~AU at any time.

\begin{acknowledgements}
We thank Julio A. Fern\'{a}ndez and Nathan Kaib for helpful reviews that greatly improved the quality of this paper.
\end{acknowledgements}

\bibliographystyle{apj}

\begin{thebibliography}{16}
\expandafter\ifx\csname natexlab\endcsname\relax\def\natexlab#1{#1}\fi

\bibitem[{{Brasser} \& {Morbidelli}(2013)}]{2013Icar..225...40B}
{Brasser}, R. \& {Morbidelli}, A. 2013, \icarus, 225, 40

\bibitem[{{Brin} \& {Mendis}(1979)}]{Brin1979}
{Brin}, G.~D. \& {Mendis}, D.~A. 1979, \apj, 229, 402

\bibitem[{{Chambers}(1999)}]{Chambers1999}
{Chambers}, J.~E. 1999, \mnras, 304, 793

\bibitem[{{Duncan} {et~al.}(1988){Duncan}, {Quinn}, \&
  {Tremaine}}]{1988ApJ...328L..69D}
{Duncan}, M., {Quinn}, T., \& {Tremaine}, S. 1988, \apjl, 328, L69

\bibitem[{{Duncan} \& {Levison}(1997)}]{1997Sci...276.1670D}
{Duncan}, M.~J. \& {Levison}, H.~F. 1997, Science, 276, 1670

\bibitem[{{Dybczy{\'n}ski} \& {Kr{\'o}likowska}(2011)}]{2011MNRAS.416...51D}
{Dybczy{\'n}ski}, P.~A. \& {Kr{\'o}likowska}, M. 2011, \mnras, 416, 51

\bibitem[{{Everhart}(1967)}]{Everhart1967}
{Everhart}, E. 1967, \aj, 72, 716

\bibitem[{{Fanale} \& {Salvail}(1984)}]{Fanale1984}
{Fanale}, F.~P. \& {Salvail}, J.~R. 1984, \icarus, 60, 476

\bibitem[{{Fern{\'a}ndez}(1981)}]{1981A&A....96...26F}
{Fern{\'a}ndez}, J.~A. 1981, \aap, 96, 26

\bibitem[{{Fern{\'a}ndez} \& {Gallardo}(1994)}]{1994A&A...281..911F}
{Fern{\'a}ndez}, J.~A. \& {Gallardo}, T. 1994, \aap, 281, 911

\bibitem[{{Fern{\'a}ndez} {et~al.}(1999){Fern{\'a}ndez}, {Tancredi}, {Rickman},
  \& {Licandro}}]{1999A&A...352..327F}
{Fern{\'a}ndez}, J.~A., {Tancredi}, G., {Rickman}, H., \& {Licandro}, J. 1999,
  \aap, 352, 327

\bibitem[{{Fouchard} {et~al.}(2013){Fouchard}, {Rickman}, {Froeschl{\'e}}, \&
  {Valsecchi}}]{2013Icar..222...20F}
{Fouchard}, M., {Rickman}, H., {Froeschl{\'e}}, C., \& {Valsecchi}, G.~B. 2013,
  \icarus, 222, 20

\bibitem[{{Gomes} {et~al.}(2008){Gomes}, {Fern Ndez}, {Gallardo}, \&
  {Brunini}}]{Gomes2008}
{Gomes}, R.~S., {Fern Ndez}, J.~A., {Gallardo}, T., \& {Brunini}, A. 2008, {The
  Scattered Disk: Origins, Dynamics, and End States}, ed. M.~A. {Barucci},
  H.~{Boehnhardt}, D.~P. {Cruikshank}, A.~{Morbidelli}, \& R.~{Dotson},
  259--273

\bibitem[{{Heisler} \& {Tremaine}(1986)}]{1986Icar...65...13H}
{Heisler}, J. \& {Tremaine}, S. 1986, \icarus, 65, 13

\bibitem[{{Hills}(1981)}]{1981AJ.....86.1730H}
{Hills}, J.~G. 1981, \aj, 86, 1730

\bibitem[{{Kaib} {et~al.}(2009){Kaib}, {Becker}, {Jones}, {Puckett}, {Bizyaev},
  {Dilday}, {Frieman}, {Oravetz}, {Pan}, {Quinn}, {Schneider}, \&
  {Watters}}]{2009ApJ...695..268K}
{Kaib}, N.~A., {Becker}, A.~C., {Jones}, R.~L., {Puckett}, A.~W., {Bizyaev},
  D., {Dilday}, B., {Frieman}, J.~A., {Oravetz}, D.~J., {Pan}, K., {Quinn}, T.,
  {Schneider}, D.~P., \& {Watters}, S. 2009, \apj, 695, 268

\bibitem[{{Kaib} {et~al.}(2011){Kaib}, {Quinn}, \&
  {Brasser}}]{Kaib2011}
{Kaib}, N.~A., {Quinn}, T., \& {Brasser}, R. 2011, \aj, 141, 3

\bibitem[{{Kr{\'o}likowska} \& {Dybczy{\'n}ski}(2010)}]{2010MNRAS.404.1886K}
{Kr{\'o}likowska}, M. \& {Dybczy{\'n}ski}, P.~A. 2010, \mnras, 404, 1886

\bibitem[{{Levison} \& {Duncan}(1994)}]{Levison1994}
{Levison}, H.~F. \& {Duncan}, M.~J. 1994, \icarus, 108, 18

\bibitem[{{Levison}(1996)}]{1996ASPC..107..173L}
{Levison}, H.~F. 1996, in Astronomical Society of the Pacific Conference
  Series, Vol. 107, Completing the Inventory of the Solar System, ed.
  T.~{Rettig} \& J.~M. {Hahn}, 173--191

\bibitem[{{Levison} \& {Duncan}(1997)}]{1997Icar..127...13L}
{Levison}, H.~F. \& {Duncan}, M.~J. 1997, \icarus, 127, 13

\bibitem[{{Levison} {et~al.}(2001){Levison}, {Dones}, \&
  {Duncan}}]{Levison2001}
{Levison}, H.~F., {Dones}, L., \& {Duncan}, M.~J. 2001, \aj, 121, 2253

\bibitem[{{Levison} {et~al.}(2006){Levison}, {Duncan}, {Dones}, \&
  {Gladman}}]{Levison2006}
{Levison}, H.~F., {Duncan}, M.~J., {Dones}, L., \& {Gladman}, B.~J. 2006,
  \icarus, 184, 619

\bibitem[{{Marsden} \& {Williams}(1993)}]{Marsden1993}
{Marsden}, B.~G. \& {Williams}, G.~V. 1993, {Catalogue of Cometary Orbits 1993.
  Eighth edition.}

\bibitem[{{Marsden} \& {Williams}(2008)}]{Marsden2008}
{Marsden}, B.~G. \& {Williams}, G.~V. 2008, Catalogue of Cometary Orbits 2008

\bibitem[{{von Neumann}(1951)}]{vonNeumann1951}
{von Neumann}, J.~V. 1950, {Nat. Bureau Standards 12}, 36-38

\bibitem[{{Nurmi} {et~al.}(2002){Nurmi}, {Valtonen}, {Zheng}, \&
  {Rickman}}]{2002MNRAS.333..835N}
{Nurmi}, P., {Valtonen}, M.~J., {Zheng}, J.~Q., \& {Rickman}, H. 2002, \mnras,
  333, 835

\bibitem[{{Oort}(1950)}]{1950BAN....11...91O}
{Oort}, J.~H. 1950, \bain, 11, 91

\bibitem[{{\"{O}pik}(1951)}]{1951PRIA...54..165O}
{\"{O}pik}, E.~J. 1951, Proc.~R.~Irish Acad.~Sect.~A, vol.~54, p.~165-199 (1951).,
  54, 165

\bibitem[{{Press} {et~al.}(1992){Press}, {Teukolsky}, {Vetterling}, \&
  {Flannery}}]{1992nrca.book.....P}
{Press}, W.~H., {Teukolsky}, S.~A., {Vetterling}, W.~T., \& {Flannery}, B.~P.
  1992, {Numerical recipes in C. The art of scientific computing}

\bibitem[{{Quinn} {et~al.}(1990){Quinn}, {Tremaine}, \&
  {Duncan}}]{1990ApJ...355..667Q}
{Quinn}, T., {Tremaine}, S., \& {Duncan}, M. 1990, \apj, 355, 667

\bibitem[{{Rickman} {et~al.}(2008){Rickman}, {Fouchard}, {Froeschl{\'e}}, \&
  {Valsecchi}}]{2008CeMDA.102..111R}
{Rickman}, H., {Fouchard}, M., {Froeschl{\'e}}, C., \& {Valsecchi}, G.~B. 2008,
  Celestial Mechanics and Dynamical Astronomy, 102, 111

\bibitem[{{Sosa} \& {Fern{\'a}ndez}(2011)}]{2011MNRAS.416..767S}
{Sosa}, A. \& {Fern{\'a}ndez}, J.~A. 2011, \mnras, 416, 767

\bibitem[{{Tisserand}(1889)}]{1889BuAsI...6..289T}
{Tisserand}, F. 1889, Bulletin Astronomique, Serie I, 6, 289
\bibitem[{{Volk} \& {Malhotra}(2008)}]{2008ApJ...687..714V}
{Volk}, K. \& {Malhotra}, R. 2008, \apj, 687, 714

\bibitem[{{Whipple}(1962)}]{Whipple1962}
{Whipple}, F.~L. 1962, \aj, 67, 1

\bibitem[{{Weissman}(1980)}]{1980A&A....85..191W}
{Weissman}, P.~R. 1980, \aap, 85, 191

\bibitem[{{Wiegert} \& {Tremaine}(1999)}]{Wiegert1999}
{Wiegert}, P. \& {Tremaine}, S. 1999, \icarus, 137, 84

\bibitem[{{Wisdom} {et~al.}(1996){Wisdom}, {Holman}, \&
  {Touma}}]{Wisdom1996}
{Wisdom}, J., {Holman}, M., \& {Touma}, J. 1996, Fields Institute
  Communications, Vol.~10, p.~217, 10, 217

\end{thebibliography}

\end{document}